\documentclass[useAMS,usenatbib,usegraphicx]{mn2e}
\usepackage{amsmath}
\title[Type IIP X-ray Emission ]{On the Lack of X-ray Bright Type IIP Supernovae}
\author[Dwarkadas]{V. V. Dwarkadas\thanks{E-mail:
    vikram@oddjob.uchicago.edu} \\ Department of Astronomy and Astrophysics, U Chicago, 5640 S Ellis Ave, Chicago, IL 60637}
  
\begin{document}
\newcommand{\vper}{\mbox{${v_{\perp}}$}}
\newcommand{\vpar}{\mbox{${v_{\parallel}}$}}
\newcommand{\uper}{\mbox{${u_{\perp}}$}}
\newcommand{\vperout}{\mbox{${{v_{\perp}}_{o}}$}}
\newcommand{\uperout}{\mbox{${{u_{\perp}}_{o}}$}}
\newcommand{\vperin}{\mbox{${{v_{\perp}}_{i}}$}}
\newcommand{\uperin}{\mbox{${{u_{\perp}}_{i}}$}}
\newcommand{\upar}{\mbox{${u_{\parallel}}$}}
\newcommand{\uparout}{\mbox{${{u_{\parallel}}_{o}}$}}
\newcommand{\vparout}{\mbox{${{v_{\parallel}}_{o}}$}}
\newcommand{\uparin}{\mbox{${{u_{\parallel}}_{i}}$}}
\newcommand{\vparin}{\mbox{${{v_{\parallel}}_{i}}$}}
\newcommand{\dout}{\mbox{${\rho}_{o}$}}
\newcommand{\din}{\mbox{${\rho}_{i}$}}
\newcommand{\da}{\mbox{${\rho}_{1}$}}
\newcommand{\mfast}{\mbox{$\dot{M}_{f}$}}
\newcommand{\mslow}{\mbox{$\dot{M}_{a}$}}
\newcommand{\beqn}{\begin{eqnarray}}
\newcommand{\eeqn}{\end{eqnarray}}
\newcommand{\be}{\begin{equation}}
\newcommand{\ee}{\end{equation}}
\newcommand{\noi}{\noindent}
\newcommand{\ftheta}{\mbox{$f(\theta)$}}
\newcommand{\gtheta}{\mbox{$g(\theta)$}}
\newcommand{\ltheta}{\mbox{$L(\theta)$}}
\newcommand{\stheta}{\mbox{$S(\theta)$}}
\newcommand{\utheta}{\mbox{$U(\theta)$}}
\newcommand{\xitheta}{\mbox{$\xi(\theta)$}}
\newcommand{\vs}{\mbox{${v_{s}}$}}
\newcommand{\ro}{\mbox{${R_{0}}$}}
\newcommand{\pa}{\mbox{${P_{1}}$}}
\newcommand{\va}{\mbox{${v_{a}}$}}
\newcommand{\vo}{\mbox{${v_{o}}$}}
\newcommand{\vp}{\mbox{${v_{p}}$}}
\newcommand{\vw}{\mbox{${v_{w}}$}}
\newcommand{\vf}{\mbox{${v_{f}}$}}
\newcommand{\lprime}{\mbox{${L^{\prime}}$}}
\newcommand{\uprime}{\mbox{${U^{\prime}}$}}
\newcommand{\sprime}{\mbox{${S^{\prime}}$}}
\newcommand{\xiprime}{\mbox{${{\xi}^{\prime}}$}}
\newcommand{\mdot}{\mbox{$\dot{M}$}}
\newcommand{\msun}{\mbox{$M_{\odot}$}}
\newcommand{\yr}{\mbox{${\rm yr}^{-1}$}}
\newcommand{\kms}{\mbox{${\rm km} \;{\rm s}^{-1}$}}
\newcommand{\lambdav}{\mbox{${\lambda}_{v}$}}
\newcommand{\lequ}{\mbox{${L_{eq}}$}}
\newcommand{\eqpratio}{\mbox{${R_{eq}/R_{p}}$}}
\newcommand{\ra}{\mbox{${r_{o}}$}}
\newcommand{\bfig}{\begin{figure}[h]}
\newcommand{\efig}{\end{figure}}
\newcommand{\tone}{\mbox{${t_{1}}$}}
\newcommand{\done}{\mbox{${{\rho}_{1}}$}}
%%%%%%%%
\newcommand{\dsn}{\mbox{${\rho}_{SN}$}}
\newcommand{\dzero}{\mbox{${\rho}_{0}$}}
\newcommand{\ve}{\mbox{${v}_{e}$}}
\newcommand{\vej}{\mbox{${v}_{ej}$}}
\newcommand{\Mch}{\mbox{${M}_{ch}$}}
\newcommand{\mej}{\mbox{${M}_{e}$}}
\newcommand{\Mst}{\mbox{${M}_{ST}$}}
\newcommand{\dam}{\mbox{${\rho}_{am}$}}
\newcommand{\Rst}{\mbox{${R}_{ST}$}}
\newcommand{\Vst}{\mbox{${V}_{ST}$}}
\newcommand{\Tst}{\mbox{${T}_{ST}$}}
\newcommand{\no}{\mbox{${n}_{0}$}}
\newcommand{\Efif}{\mbox{${E}_{51}$}}
\newcommand{\rsh}{\mbox{${R}_{sh}$}}
\newcommand{\msh}{\mbox{${M}_{sh}$}}
\newcommand{\vsh}{\mbox{${V}_{sh}$}}
\newcommand{\vrev}{\mbox{${v}_{rev}$}}
\newcommand{\rpr}{\mbox{${R}^{\prime}$}}
\newcommand{\mpr}{\mbox{${M}^{\prime}$}}
\newcommand{\vpr}{\mbox{${V}^{\prime}$}}
\newcommand{\tpr}{\mbox{${t}^{\prime}$}}
\newcommand{\cone}{\mbox{${c}_{1}$}}
\newcommand{\ctwo}{\mbox{${c}_{2}$}}
\newcommand{\cthree}{\mbox{${c}_{3}$}}
\newcommand{\cfour}{\mbox{${c}_{4}$}}
\newcommand{\Te}{\mbox{${T}_{e}$}}
\newcommand{\Ti}{\mbox{${T}_{i}$}}
\newcommand{\Ha}{\mbox{${H}_{\alpha}$}}
\newcommand{\Rprime}{\mbox{${R}^{\prime}$}}
\newcommand{\Vprime}{\mbox{${V}^{\prime}$}}
\newcommand{\Tprime}{\mbox{${T}^{\prime}$}}
\newcommand{\Mprime}{\mbox{${M}^{\prime}$}}
\newcommand{\rprime}{\mbox{${r}^{\prime}$}}
\newcommand{\rfprime}{\mbox{${r}_f^{\prime}$}}
\newcommand{\vprime}{\mbox{${v}^{\prime}$}}
\newcommand{\tprime}{\mbox{${t}^{\prime}$}}
\newcommand{\mprime}{\mbox{${m}^{\prime}$}}
\newcommand{\Me}{\mbox{${M}_{e}$}}
\newcommand{\nh}{\mbox{${n}_{H}$}}
\newcommand{\rr}{\mbox{${R}_{2}$}}
\newcommand{\rf}{\mbox{${R}_{1}$}}
\newcommand{\vtwo}{\mbox{${V}_{2}$}}
\newcommand{\vout}{\mbox{${V}_{1}$}}
\newcommand{\dshell}{\mbox{${{\rho}_{sh}}$}}
\newcommand{\dwind}{\mbox{${{\rho}_{w}}$}}
\newcommand{\dslow}{\mbox{${{\rho}_{s}}$}}
\newcommand{\dfast}{\mbox{${{\rho}_{f}}$}}
\newcommand{\vfast}{\mbox{${v}_{f}$}}
\newcommand{\vslow}{\mbox{${v}_{s}$}}
\newcommand{\cc}{\mbox{${\rm cm}^{-3}$}}
\newcommand{\apj}{\mbox{ApJ}}
\newcommand{\apjl}{\mbox{ApJL}}
\newcommand{\apjs}{\mbox{ApJS}}
\newcommand{\aj}{\mbox{AJ}}
\newcommand{\araa}{\mbox{ARAA}}
\newcommand{\nat}{\mbox{Nature}}
\newcommand{\aap}{\mbox{AA}}
\newcommand{\gca}{\mbox{GeCoA}}
\newcommand{\pasp}{\mbox{PASP}}
\newcommand{\mnras}{\mbox{MNRAS}}
\newcommand{\apss}{\mbox{ApSS}}

\date{}

\pagerange{\pageref{firstpage}--\pageref{lastpage}} \pubyear{2014}

\maketitle

\label{firstpage}

\begin{abstract}
Type IIP Supernovae (SNe) are expected to arise from Red Supergiant
stars (RSGs). These stars have observed mass-loss rates that span more
than two orders of magnitude, from $< 10^{-6} \msun$ yr$^{-1}$ to
almost $ 10^{-4} \msun$ yr$^{-1}$. Thermal bremsstrahlung X-ray
emission from at least some IIP's should reflect the larger end of the
high mass-loss rates. Strangely, no IIP SNe are seen where the X-ray
luminosity is large enough to suggest mass-loss rates greater than
about $ 10^{-5} \msun$ yr$^{-1}$. We investigate if this could be due
to absorption of the X-ray emission. After carefully studying all the
various aspects, we conclude that absorption would not be large enough
to prevent us from having detected X-ray emission from high mass-loss
rate IIP's. This { leads us to the conclusion} that there may be an
upper limit of $\sim 10^{-5} \msun$ yr$^{-1}$ to the mass-loss rate of
Type IIP progenitors, and therefore to the luminosity of RSGs that
explode to form Type IIPs. This is turn suggests an upper limit of
$\la 19 \msun$ for the progenitor mass of a Type IIP SN. This limit is
close to that obtained by direct detection of IIP progenitors, as well
as that suggested by recent stellar evolution calculations. Although
the statistics need to be improved, many current indicators support
the notion that RSGs above $\sim 19 \msun$ do not explode to form Type
IIP SNe.

\end{abstract}

\begin{keywords}
Shock waves; circumstellar matter; stars: massive; supernovae:
general; stars: winds, outflows; X-rays: ISM
\end{keywords}
%% main text
\section{Introduction}
\label{sec:intro}

It is generally assumed that supernovae (SNe) evolve in the
circumstellar wind medium carved out by their progenitor stars
\citep{chevalier82a}.  The resulting emission from SNe, especially in
the radio and X-ray, is due to the interaction of the SN shock wave
with this circumstellar medium \citep{chevalier82b}. The magnitude of
the emission, especially the thermal X-ray emission, is a function of
the circumstellar medium (CSM) parameters, in particular the density
profile and structure of the circumstellar medium. This can be used to
our advantage, so that the X-ray light curves of SNe can be used to
uncover the density profile of the CSM \citep[see for
  example][]{ddb10}.

Stellar evolution theory posits that Type IIP SNe arise from Red
Supergiant progenitors \citep{schalleretal92, langer93, maeder2000,
  langer12}, which have low velocity winds of order 10 km s$^{-1}$,
with mass-loss rates that span from 10$^{-6}$ to 10$^{-4} \msun$
yr$^{-1}$. All clearly identified progenitors of Type IIP SNe appear
to be RSGs \citep{smartt09}. Due to their low velocities and high
mass-loss rates, the wind densities (${\rho}_w \propto \dot{M}/(4 \pi
r^2 v_w)$) around RSG stars are expected to be some of the highest
around SN progenitor stars. In particular the wind densities around
Type IIP SNe are expected to be much higher than those around Type
Ib/c SNe, whose progenitors, thought to be Wolf-Rayet (W-R) stars,
have much higher wind velocities, of order 1000 km s$^{-1}$, and thus
wind densities that are proportionally lower. Thermal bremsstrahlung
X-ray emission from SNe, due to circumstellar interaction
\citep{chevalier82b}, is proportional to the density
squared. Consequently, { if the X-ray emission is due to thermal
  bremsstrahlung}, Type IIP SNe as a group would be expected to have
some of the highest x-ray luminosities. The observations however
indicate exactly the opposite, with IIP's having the lowest X-ray
luminosities of all observed SNe (Figure 1).

The goal of this paper is to investigate the lack of X-ray bright Type
IIP SNe, probe whether this is a real problem in the first place, and
if so study the implications of this fact. In \S 2 we discuss the
observed wind parameters of RSG stars. \S 3 displays the observed
X-ray lightcurves of young SNe. \S 4 investigates whether absorption
of the X-ray emission would prevent us from detecting type IIP SNe
with high mass-loss rates. \S 5 discusses the implications of the
results, and suggests that there may be a maximum mass above which RSGs
do not explode to form IIp's. \S 6 summarizes the research and
elucidates the important conclusions.

\section{RSG Mass-loss Parameters}
\label{sec:rsgmdot}

Several empirical prescriptions have been proposed in the literature
to express the RSG mass-loss rate as a function of the stellar
parameters \citep{reimers75, deJager88, ndj90, vdv98, sbc99,
  vanloonetal05}. While they vary somewhat in their precise
formulation, they all indicate that the mass-loss rate is proportional
to some power of the RSG luminosity, with more luminous RSGs having
higher mass-loss rates.  Since the luminosity of the star increases
with increasing stellar mass, it is clear that the mass-loss rate
increases with increasing mass.

\citet{mj11} have compared the mass-loss rate prescriptions with
observationally calculated mass-loss rates of RSGs. Of the 8 RSGs
which have a mass-loss rate measured from circumstellar gas
observations, 3 have mass-loss rates greater than 10$^{-5} \msun$
yr$^{-1}$, with at least one of these (depending on what measurement
method is used) having a mass-loss rate greater than 10$^{-4} \msun$
yr$^{-1}$. For another set of 39 RSGs, whose mass-loss rates were
derived from the infrared excess at 60 $\mu$m, 4 were found to have
mass-loss rates higher than 10$^{-5} \msun$ yr$^{-1}$. For RSGs in the
LMC, there was considerably more scatter among different methods used
to estimate the mass-loss rates, with the most conservative one (using
{\it Spitzer} data), having mass loss rates generally below 10$^{-5}
\msun$ yr$^{-1}$ with one exception, whereas those using IRAS fluxes
generally had mass-loss rates exceeding 10$^{-5} \msun$ yr$^{-1}$.

The terminal velocities of RSG winds are hard to measure, but are low,
and less than the escape velocities from the star. Although the
velocities are generally taken to be around 10 km s$^{-1}$, there is
some variation in the velocities. \citet{josselinetal00} suggest
velocities of 25 km s$^{-1}$ for M supergiants, following the
observational work of \citet{jura86} and \citet{km85}. \citet{mj11}
find velocities for RSG winds generally above 10 km s$^{-1}$, and
stretching all the way up to 50 km s$^{-1}$. Furthermore, there is
some indication of a variation with luminosity, and therefore with
mass-loss rate, with those having higher mass-loss rates also having
higher velocities, although this relationship isn't
confirmed. Velocities for RSGs in the LMC follow the same trend, but
appear to be about 30\% lower.

There understandably exists considerable confusion over observed
mass-loss parameters of RSGs, given the various measurement methods
used, and the fact that RSG properties are difficult to
measure. Nevertheless, there seems to be general agreement that there
exist at least a few RSGs with {\em measured} mass-loss rates
exceeding 10$^{-5} \msun$ yr$^{-1}$. \citet{humphreys07} mentions a
few post-RSGs, or cool hypergiants, with mass-loss rates exceeding
10$^{-4} \msun$ yr$^{-1}$, at least for short periods of time. {\em
  Theoretical considerations} imply that as the luminosity increases,
the mass-loss rate should increase. RSGs with luminosities $> 3 \times
10^5 L_{\odot}$ are observed, corresponding to large progenitor masses
and therefore larger mass-loss rates. It is then problematic that not
a single Type IIP SN has been seen with a high X-ray luminosity (\S
\ref{sec:iipxray}), indicating a mass-loss rate $>$ 10$^{-5} \msun$
yr$^{-1}$. This cannot be purely a selection problem, since if the RSG
winds are observable there is no reason why the brighter Type IIP SNe
should not be observable.

\section{X-Ray emission from Type IIP SNe}
\label{sec:iipxray}

\begin{figure*}
\includegraphics[scale=0.7, angle=90]{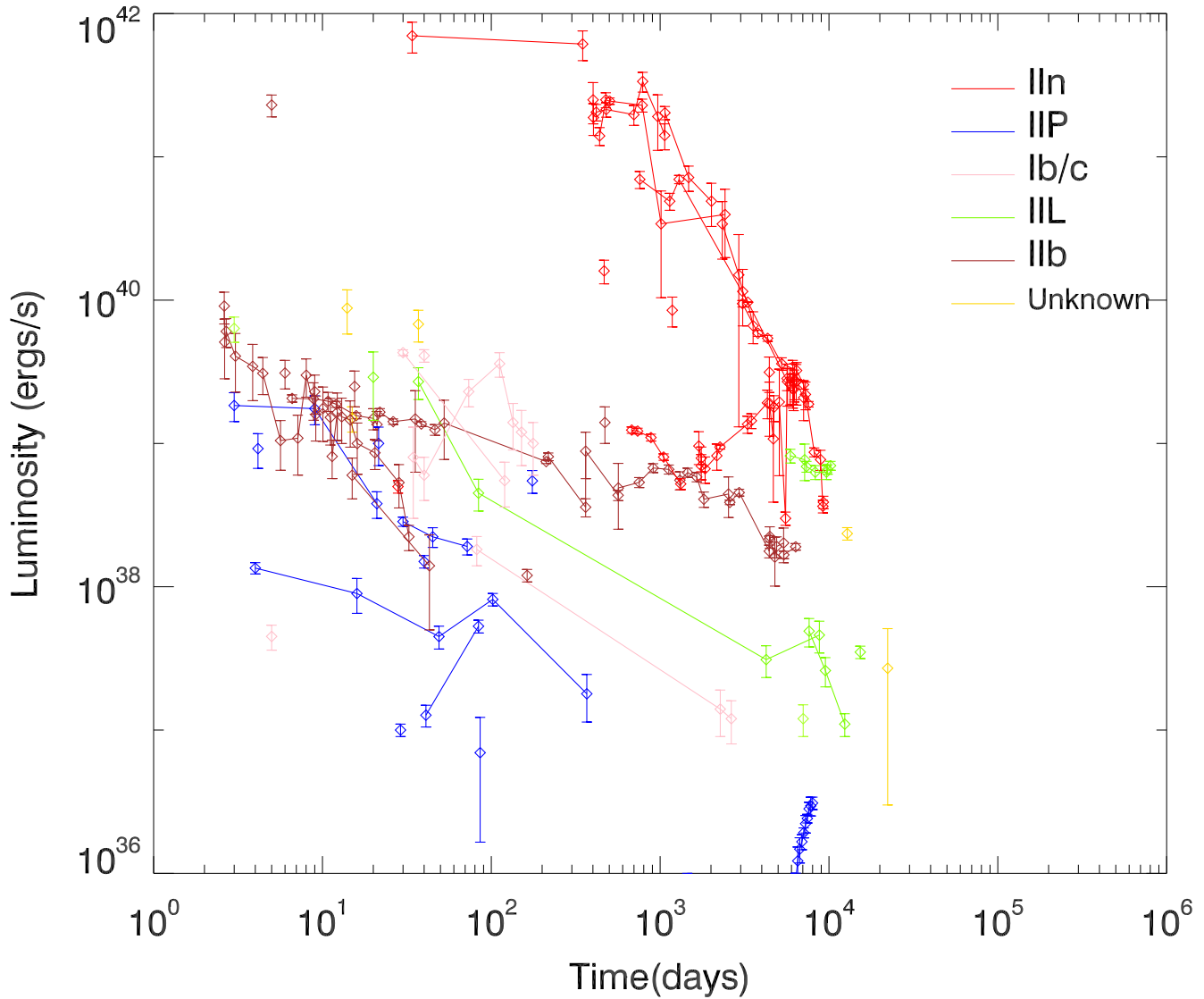}
\caption{The light curves of most published SNe, by type. This figure
  is adapted from a similar figure first published in \citet{dg12},
  with many more SNe added. It is clearly seen that the light curves
  of Type IIP SNe generally lie below a luminosity of 10$^{39}$ ergs
  s$^{-1}$, almost three orders of magnitude below the brightest
  observed SNe. Note: GRB SNe are not included in this list.
\label{fig:lctype}}
\end{figure*}

Figure \ref{fig:lctype} shows the X-ray emission from young SNe,
grouped by type. As can be seen, Type IIP SNe have the lowest levels
of X-ray emission\footnote{We note that 1994W was classified as Type
  IIP by \citet{scl98} following the photometry of \citet{t95}. The
  high X-ray luminosity would have made it probably the brightest
  observed IIP in X-rays. However it has subsequently been cited as
  Type IIn by most authors. It is possible that it belongs to the
  group of SNe whose classification evolves with SN evolution. However
  its properties, and X-ray emission, appear to place it amongst the
  IIns. In this paper we have classified it as a IIn}. Bremsstrahlung
emission is proportional to the square of the density. If, in the
simplest approximation, this plot is looked on as representative of
the ambient density in which the SNe evolve, with X-ray emission due
to thermal bremsstrahlung, then it would appear as if Type IIP's
evolve in a medium with the lowest density, and Type IIns in media
with the highest density. While it is very possible that many IIns
evolve in media with very high densities, it seems strange that IIPs,
at least a few of which should have very high density media around
them, appear to have the lowest luminosities as a group.

\citet{cf03} give the free-free X-ray luminosity from the SN shocks:

\be
L_x \approx 3. \times 10^{39}\;g_{ff}\;C_n\;\left(\frac{{\dot{M}}_{-5}}{v_{w1}}\right)^2\,t_{10}^{-1}
\label{eq:lx}
\ee

\noi where $g_{ff}$ is the gaunt factor, of order unity, $C_n = 1$ for
the circumstellar medium shock, and $C_n = (n-3)(n-4)^2/[4(n-2)]$ for
the reverse shock, ${\dot{M}}_{-5}$ is the mass loss rate in units of
10$^{-5} \msun$ yr$^{-1}$, $v_{w1}$ is the wind velocity in units of
10 km s$^{-1}$, and $t_{10}$ is time in units of 10 days. This assumes
electron-ion equilibration, and a freely-expanding wind with constant
mass-loss parameters. It is clearly seen that, given the luminosities
in Fig 1 all being below 3 $\times 10^{39}$, all the observed X-ray
epochs for type IIP SNe suggest mass-loss rates below 10$^{-5} \msun$
yr$^{-1}$ for a wind velocity of 10 km s$^{-1}$, {if the emission
  is due to thermal bremsstrahlung}. If the wind velocity is a factor
of 2-3 higher, then the mass-loss rate is correspondingly higher, and
for the most luminous cases may exceed ${\dot{M}}_{-5} =1$.

In Table 1 we show the mass-loss rates, and progenitor masses, of Type
IIP SNe computed by various authors. These are calculated from radio,
x-ray and optical data. We also show the corresponding calculated
progenitor mass for the SN, which are calculated by a variety of
means, including direct detection of progenitors in pre-explosion
images \citep{smarttetal09}, spectroscopic modelling, and hydrodynamic
modelling. {The method is identified in column 4}.

\begin{table*}
\caption{The progenitor masses, and mass-loss rates, deduced for Type
  IIP SNe in the literature. References: (1) \citet[][and references
    within]{cfn06} (2) \citet{royetal11} (3) \citet{ucb10} (4)
  \citet{moriyaetal11} (5) \citet{misraetal07} (6)
  \citet{jerkstrandetal12} (7) \citet{chakrabortietal12} (8)
  \citet{chakrabortietal13} (9) \citet{moriyaetal12a} (10)
  \citet{tomasellaetal13} (11) \citet{uc13} (12) \citet{uc08} (13)
  \citet{uc09} (14) \citet{ucp07} (15) \citet{uc11} (16)
  \citet{boseetal13} (17) \citet{inserraetal12} (18)
  \citet{inserraetal11} (19) \citet{vandyketal12} (20)
  \citet{maundetal13a} (21) \citet{mrm14} (22) \citet{mmre14} (23)
  \citet{smarttetal09} (24) \citet{vandyketal12b} }
\begin{tabular}{|l|c|c|c|c|} \hline
 SN & Progenitor Mass ($\msun$) & Mass-Loss Rate (10$^{-6} \msun$ yr$^{-1}$) & Mass (or Mass-loss Rate) Method & Reference \\  \hline
1999an & $<$ 18 & $\ldots$ & Optical progenitor identification & 23 \\
1999br & $<$ 15 & $\ldots$ &  Optical progenitor identification & 23 \\
1999em & $<$ 15 & 5 & radio synchrotron; Optical progenitor identification & 1, 23 \\ 
1999ev & $16^{+6}_{-4}$ & $\ldots$ & Optical progenitor identification & 23 \\
1999gi & $<$ 14 & $\ldots$ & Optical progenitor identification &  23 \\
2000cb & 26.3 $\pm$ 2 & $\ldots$ & Hydrodynamic model & 15 \\
2001du & $<$ 15 & $\ldots$ & Optical progenitor identification &  23 \\
2002hh & $<18$ & 7 & radio synchrotron; Optical progenitor identification & 1, 23 \\ 
2003gd & 8.4 $\pm$ 2 & $\ldots$ & Optical progenitor identification & 21 \\
2003gd & $7^{+6}_{-2}$ & $\ldots$ & Optical progenitor identification & 23 \\
2003ie & $<$ 25 & $\ldots$ & Optical progenitor identification & 23 \\
2003Z  & 14.4 -17.4 & $<$ 0.16 & Hydrodynamic model & 14 \\
2004A  & 12 $\pm$ 2.1 & $\ldots$ & Optical progenitor identification & 21 \\
2004A  & 7$^{+6}_{-2}$ & $\ldots$ & Optical progenitor identification & 23 \\
2004am & $12^{+7}_{-3}$ & $\ldots$ & Optical progenitor identification & 23 \\
2004dg & $<$ 12 & $\ldots$ & Optical progenitor identification & 23 \\
2004dj & 15, $\sim$ 12 & 2-3 & radio synchrotron & 1 \\
2004dj & 15 & 0.32 $\pm$ 0.11 & X-ray, radio; observations and theory & 7 \\
2004dj & 15 $\pm$ 3 & $\ldots$ & Optical progenitor identification & 23 \\ 
2004et & 15$^{+5}_{-2}$ & 9-10 & radio synchrotron & 1 \\ 
2004et & $\sim$ 20 & $\sim$ 2 & X-ray luminosity & 5 \\
2004et & 27 $\pm$ 2 & $\ldots$ & Hydrodynamic model & 13 \\
2004et & $<$ 15 & $\ldots$ & Late-time spectral modelling & 6 \\
2004et & 9$^{+5}_{-1}$ & $\ldots$ & Optical - direct progenitor identification & 23 \\
2005cs & 18.2 $\pm$ 1 & $\ldots$ & Hydrodynamic model &  12 \\
2005cs & 9.5$_{-2.2}^{+3.4}$ & $\ldots$ & Optical progenitor identification & 21 \\
2005cs & 7$^{+3}_{-1} $ & $\ldots$ & Optical progenitor identification & 23 \\
2006bc & $<$ 12 & $\ldots$ & Optical progenitor identification & 23 \\
2006my & 9.8 $\pm$ 1.7 & $\ldots$ & Optical progenitor identification & 21 \\
2006my & $<13$ & $\ldots$ & Optical progenitor identification & 23 \\
2006ov & $<10$ & $\ldots$ & Optical progenitor identification & 23 \\
2007aa & $<12$ & $\ldots$ & Optical progenitor identification & 23 \\
2007od & 9.7-11 & $\ldots$ & Modelling bolometric light curve & 18 \\
2008bk & 8-8.5 & $\ldots$ & Evolutionary Models & 19 \\
2008bk & 12.9 $\pm$ 1.7 & $\ldots$ & Optical progenitor identification & 22 \\
2008bk & 9$^{+4}_{-1}$ & $\ldots$ & Optical progenitor identification & 23 \\
2008in & $<$ 20; 15.5 $\pm$ 2.2 & $<$ 5 & Hydrodynamical model & 2, 11 \\
2009bw & 11-15 & $\sim$ 1 ($<$ 100) & Modelling light curve & 17 \\
2009kf & $\sim$ 36 & $< 90$ & Hydrodynamic Modelling & 3 \\
2009kf & $\ldots$ & $> 100$, 10000 & radiation hydrodynamics & 4, 9 \\
2011ja & $> $ 16 & 0.27 $\pm$ 0.05; 7.5 $\pm$ 1.5 & X-ray, radio fitting & 8 \\
2012A  & 10-15 & $\ldots$ & Optical progenitor; Hydrodynamic Modelling & 10 \\ 
2012aw & 14-15 & $\ldots$ & Observations; Hydrodynamical Modelling & 16 \\
2012aw & 15-20 & $\ldots$ & Evolutionary tracks & 24  \\
2012ec & 14-22 & $\ldots$ & Optical - photometric and Spectroscopic & 20 \\
\end{tabular}
\end{table*}

Data on some SNe in Table 1 is expanded over multiple rows, where
multiple groups using different techniques calculated mass-loss rates
or progenitor masses, using data over a multitude of wavelengths. In
all but one case, the calculated mass-loss rate is lower than 10$^{-5}
\msun$ yr$^{-1}$. In many cases where a mass-loss rate is not
specified, the inferred progenitor mass would suggest a mass-loss rate
below 10$^{-5} \msun$ yr$^{-1}$.

Progenitor masses calculated from hydrodynamical modelling appear to
be systematically higher than those calculated from the data, either
spectra or light-curves. A major exception to the low mass-loss rate
appears to be the ultraviolet-bright SN 2009kf
\citep{botticellaetal10}, whose early ultraviolet lightcurve appears
to indicate either a very large energy and high-mass progenitor
\citep{ucb10}, or an exceptionally high mass-loss rate
\citep{moriyaetal11}.

\section{Suggested Reasons for the lack of X-ray Bright IIP's}

As mentioned above, there are clear indications that RSGs with high
mass-loss rates exist. These are expectedly fewer in number,
consistent with a steep initial mass-function that is strongly biased
towards lower-mass stars. It is also clear that IIPs arise from
RSGs. Almost all SNe whose progenitors are reasonably well confirmed
appear to be Type IIP SNe, and their progenitors appear to be RSGs
\citep{smartt09, smarttetal09}.  However none of the progenitors
appear to have an initial mass exceeding 16.5 $\pm$ 1.5 $\msun$. The
lack of high mass RSG progenitors of Type IIP SNe has been touched on
by \citet{smarttetal09}, who refer to it as the ``red supergiant
problem''.

\subsection{Absorption of the X-ray Emission}

The majority of observations of Type IIP SNe in X-rays
(\ref{fig:lctype}) appear to be at an age of $<$ 100
days. \citet{cfn06} suggest that the early emission from Type IIP SNe
may be due to inverse Compton scattering of the photospheric photons.
However, {if the mass-loss rate of the RSG were higher}, thermal
bremsstrahlung would be expected to dominate. \citet{cfn06} also
suggest that emission from the reverse shock would dominate, unless
the reverse shock were to be radiative. For mass-loss rates $> 10^{-5}
\msun$ yr$^{-1}$, and an age $< 100$ days, the reverse shock in IIP's
is expected to be radiative \citep{flc96, cfn06}, and a cool shell is
expected to form that will absorb almost all the emission from the
reverse shock. Thus for higher mass-loss rates, one would expect the
emission to arise from the forward shocked circumstellar wind material
in Type IIP SNe.

We explore if emission from the circumstellar shock in high luminosity
IIps could be absorbed by the forward-shocked material, as has been
suggested by some authors \citep{moriyaetal12a}. The velocity of the
forward shock can be written as \citep{cfn06}:

\begin{align}
\label{eq:vel}
V_f = & 1.6 \times 10^4 \left[\frac{{\dot{M}}_{-6}}{v_{w1}}\right]^{-0.1}\;E_{51}^{0.45}\;\left[\frac{M_{ej}}{10 \,\msun}\right]^{-0.345} \nonumber \\
 & \times \,\left[\frac{t}{10\; {\rm days}}\right]^{-0.10} {\rm km\; s^{-1}}
\end{align}

\noi where $v_{w, 1}$ is the wind velocity in units of 10 km s$^{-1}$,
and ${\dot{M}}_{-6}$ is the wind mass-loss rate in units of 10$^{-6}
\msun$ yr$^{-1}$.  This is for a specific value of the ejecta density
profile $n=11.73$, and will vary slightly for different profiles. Note
that the variation {of the forward shock velocity} with mass-loss
rate (assuming a constant { wind} velocity) is small - modifying
the mass-loss rate by a factor of 100 decreases the {forward
  shock} velocity to 0.63 times its original value. This equation can
be used to approximate the velocity evolution.

Since the ionization structure of the ambient medium plays a
significant role in determining the photo-electric absorption, the
ionization structure of the surrounding wind material needs to be
investigated. We first consider the ionization from the SN explosion
and resulting UV flash. For a medium consisting of pure H,
\citet{lf96} estimate that the burst can ionize anywhere from 0.5 to 2
$\msun$ of material. {Although these models are for Blue (and not
  Red) supergiant progenitors, they provide an approximate estimate}
that the SN explosion should ionize around 0.5 $\msun$. This will then
ionize the RSG wind out to a radius $R_i$ where

\be
R_{i} = 1.6 \times 10^{19} \, M_{i, 0.5}\,v_{w, 1}\,{\dot{M}}_{-6}^{-1}\; {\rm cm}
\ee

\noi where $M_{i, 0.5}$ is the total mass that can be fully ionized by
the explosion, in units of 0.5 $\msun$. It can be seen that even for
mass-loss rates $\sim 10^{-4} \msun$ yr$^{-1}$, the wind material can
be ionized out to 10$^{17}$ cm.

However the densities close to the star are extremely high, and
recombination proceeds rapidly. The recombination time can be
estimated as

\be
t_{rec} \sim \frac{3 \times 10^{12}}{n_e} \sim 1.2 \times 10^6\, r_{15}^2\,v_{w,1}\,{\dot{M}}_{-6}^{-1}\; {\rm s}
\label{eq:rec}
\ee

\noi where $r_{15}$ is the radius in units of 10$^{15}$ cm.

Using equations \ref{eq:vel} and \ref{eq:rec} suggests that for
${\dot{M}}_{-6} =1$, the recombination time will be larger than the SN
age after about 30 days. Therefore the wind from that radius out
should be ionized, and optically thin to the thermal X-ray flux.

This is consistent with observational data. \citet{misraetal07} find
the mass-loss rate for SN 2004et to be ${\dot{M}}_{-6} \la 2$ from the
X-ray data. For 3 {\it Chandra} observations, taken at 30, 45 and 72
days after explosion, the column density is consistent with the
Galactic value, with no additional column around the star, suggesting
the presence of an ionized wind that does not add to the
absorption. Our own analysis of the data finds results consistent with
the fluxes obtained by \citet{misraetal07} within the error bars. We
also find that if we set the minimum column to be the Galactic value
\citep{dl90, kalberlaetal05}, the best fit to the spectrum is always
obtained at this minimum value.

For higher mass-loss rates, the recombination time decreases
appropriately. For ${\dot{M}}_{-6} =10$, the recombination time is
smaller than the SN shock flow time for about 5 months, whereas for
${\dot{M}}_{-6} =100$, the recombination time is less than the SN
shock travel time for about 4 years. We can consider that for all the
early observations, at mass-loss rates $> {\dot{M}}_{-6} =10$, the
medium no longer shows the effects of the initial ionizing radiation. 

{However}, the X-ray emission itself can photo-ionize the
medium. The importance of this mechanism depends on the value of the
ionization parameter $\chi = L_x / (n\,r^2)$ \citep{km82}. Assuming
the X-ray luminosity to be given by equation \ref{eq:lx}, we can write

\be
\chi = 120\;{\dot{M}}_{-5}\; v_{w1}^{-1}\;t_{10\; {\rm days}}^{-1}
\ee

\citet{ci12} have studied the effects of $\chi$ on the ionization. For
$\chi \ga 100$, the elements C, N, O are completely ionized; however
ionization of the heavier elements requires $\chi \ga 1000$. The
results are also somewhat dependent on the temperature of the
radiation field. For ${\dot{M}}_{-5} \sim 10 $ most of the elements
will be highly ionized, and the medium may be considered mostly
ionized in the first couple of weeks. By about 100 days though the
heavier elements will not be fully ionized, and thus some absorption
by Si and Fe could be expected. For ${\dot{M}}_{-5} \sim 1 $, C, N and
O are ionized in the first few days, but only low ionization stages
are present after the first few weeks.

Unless the medium is fully ionized, which is true only during the
first couple of weeks for the largest mass-loss rates considered here,
some absorption of the X-ray emission is expected. Although it would
require a time-dependent ionization calculation to compute accurately,
some estimates can be made. In the wavelength range between 0.2-4 KeV,
where we expect much of the X-ray absorption to take place, the most
important contributions to the opacity arise from K-shell absorption
by carbon, nitrogen and oxygen. \citet{fransson82} estimates the
optical depth at an energy E to be

\be
\label{eq:tau}
\tau (E) \sim 43 {\dot{M}}_{-4} \,v_{w,6}^{-1}\,r_{15}^{-1}\,E_{keV}^{-8/3}
\ee

\noi \citet{fransson82} indicates that the expression is correct to
within a factor of two independent of the ionization state, as long as
the atoms are not completely stripped of electrons, i.e. that the
medium is not totally ionized. 

In the first 10 days we would not expect the SN shock to have
travelled more than 10$^{15}$ cm, and in the first 100 days one would
expect the shock radius to be around 10$^{16}$ cm. Thus, from equation
\ref{eq:tau}, for all mass-loss rates $ \ga 2 \times 10^{-6} \msun$
yr$^{-1}$, the optical depth at 1 keV will be larger than 1. For a
velocity of 10$^4$ km s$^{-1}$, we find that the distance traveled in
one month will be 2.6 $\times 10^{15}$ cm. For a mass-loss rate of
${\dot{M}}_{-5} \sim 1 $, and a somewhat higher wind velocity of 20 km
s$^{-1}$ (see \S \ref{sec:rsgmdot}), the optical depth has dropped
below unity already, and continues to fall as the SN shock expands
outwards. Thus at times later than about a month after explosion, and
for a mass-loss rate around 10$^{-5} \msun$ yr$^{-1}$, we would expect
that while there is some absorption, the total flux is large enough to
be observable. At higher mass-loss rates, while the optical depth is
still high, note that the luminosity is increasing as the square of
the mass-loss rate. Thus for a mass-loss rate of ${\dot{M}}_{-5} = 5$
and wind velocity 20 km s$^{-1}$, we find that the luminosity
increases by over a factor of 6 compared to the previous case. At
about 30 days, the optical depth is around 4, but the total flux, for
a SN at 10 Mpc, is about 3 $\times 10^{-14}$ ergs s$^{-1}$ cm$^{-2}$,
using equation \ref{eq:lx}. This is about half that noted for the last
2004et observation, and detectable. At later times the luminosity
decreases but the optical depth also decreases.

At the highest RSG mass-loss rates of 10$^{-4} \msun$ yr$^{-1}$, the
medium can be considered totally ionized for the first two weeks due
to photo-ionization from the shock. Subsequently, for the first 100
days C, N, and O will still be almost fully ionized, although heavier
species will not, and some absorption will undoubtedly occur. But
following the argument above, particularly if the wind velocity
exceeds 30 km s$^{-1}$ at the highest mass-loss rates as observations
suggest, the optical depth will exceed unity, but the X-ray flux for a
SN within 10 Mpc will be in the range of a few times 10$^{-14}$ergs
s$^{-1}$ cm$^{-2}$. Furthermore, assuming that C, N and O are fully
ionized will reduce the photo-absorption and the optical depth to
values below that quoted above, and thus increase the observed flux.

We can examine this from another perspective. Many of the type IIP SNe
that are seen have their X-ray luminosity, at least early on,
dominated by IC processes, which exceeds the thermal
bremsstrahlung. However, other SN types with mass-loss rates around
10$^{-5} \msun$ yr$^{-1}$ or somewhat less are clearly seen (Figure
1), so there should be no reason for IIPs with similar mass-loss rates
to go undetected. Furthermore, as the emission goes as density
squared, the luminosity will increase by a factor of 100 for an
increase of a factor of 10 in mass-loss (equation \ref{eq:lx}). If the
luminosity is 100 times higher for the highest mass-loss rate ones,
even an optical depth of 4 will return about the same flux as the
lower mass-loss rate SN (assuming the same distance of course) and
thus should be detectable. Therefore, given that other SNe are
detectable, and thermal emission from some low density type IIPs is
seen, it becomes hard to justify that not a single high mass-loss rate
one has been detected.

If all the emission were to fall in the {\it Chandra} or {\it XMM}
band, it would be detectable with a 50 ks observation.  Next we
investigate what fraction of the emission from the forward shock would
fall in the 0.5-10 keV range over which current X-ray satellites
work. For a shock velocity 10$^4$ km s$^{-1}$, the post-shock
temperature will be about 10$^9$ K. If we assume that electrons are
heated purely by Coulomb collisions, the time taken for electrons and
ions to reach temperature equilibrium is:

\be
t_{eq} = 5 \times 10^{5} \; \frac{T_{e9}^{3/2}}{n_{e9}} \; {\rm s}
\ee

\noi where $T_{e9}$ is the electron temperature in units of 10$^9$K,
and $n_{e9}$ is the electron density in units of 10$^9$ cm$^{-3}$. Given that 

\be
n_{e9} = 0.1 {\dot{M}}_{-5}\,v_{w,1}^{-1}\,r_{15}^{-2}
\ee

we have 

\be
t_{eq} = 5 \times 10^{6} \; T_{e9}^{3/2}\, \dot{M}_{-5}^{-1}\,v_{w,1}\,r_{15}^{2}\; {\rm s}
\ee

\noi Therefore, except at the highest mass-loss rates ($> 10^{-4}
\msun$ yr$^{-1}$), and large SN velocities, i.e in the first couple of
weeks, the equilibration time is larger than the age of the SN. Thus
the electrons will not in general attain temperature equilibrium with
the ions, and the electron temperature will be lower than the ion
temperature. Collisionless plasma processes may play a role, but at
these high densities will not be expected to dominate. This is
consistent with the results of \citet{ci12} as expressed in their
Figure 1.

{In equation (1) we assumed that the luminosity was due to
  electron-ion equilibration. If this is not achieved, as outlined
  above, and } the electron temperature is lower than the post-shock
temperature, what fraction of the flux falls into the {\it Chandra}
1-10 keV band? This depends on many factors such as the ratio of
electron-ion temperature, steepness of the ejecta density profile, and
density of the medium, and requires a detailed exploration of the
parameter space. However, the basic idea has been summarized in the
calculations by \citet{flc96} (see their figures 8 and 10). The X-ray
flux in the 1-10 keV band in their calculations for SN 1993J, with
mass-loss rates comparable to those we are assuming here, is a factor
of 5-10 lower than the total X-ray flux. Note that these calculations
assume that $T_e = T_{coul}$, {and it is possible that plasma
  processes could make the electron temperature greater than the
  Coulomb temperature (although still lower than the ion
  temperature)}. Given the flux that we computed earlier, a factor of
a few lower in the first 100 days makes the SN harder to detect, but
it is still detectable with long exposures of 50-100 ks. As expected,
over time the post-shock temperature decreases and more and more of
the X-ray flux falls into the {\it Chandra} and {\it XMM} bands. After
a few hundred days or so, most of the flux from the SN shock lies in
the 1-10 keV band.

So far we have considered the emission mainly within 100 days after
explosion. As is evident from the above discussion, while the
luminosity will be decreasing with time, the temperature will also be
decreasing, thus moving a larger fraction of the emitted flux into the
{\it Chandra} wavebands. The medium further out will have a much lower
opacity, and thus most of the X-ray emission will manage to escape
unabsorbed. Therefore at later times (several months to a year) it
should also be possible to detect X-ray emission from high mass-loss
rate Type IIPs.

Furthermore, {as the SN shock advances in radius}, and the density
in the wind decreases, the reverse shock will no longer be radiative,
a dense shell will not form that absorbs most of the reverse-shocked
emission, and the X-ray emission from the reverse shock will begin to
contribute. After a 100 days, the column density behind the reverse
shock, for a mass-loss rate of $> 10^{-5} \msun$ yr$^{-1}$, is a few
times 10$^{21}$ cm$^{-2}$ \citep{cfn06}. Thus most of the emission
below 1 keV will be absorbed, but we would still expect emission $> 1$
keV to be visible.

{It could be argued that late-time X-ray emission from a IIP SN
  would not be detected, because there would be no reason to observe
  the SN if early-time emission was not detected. While this is true,
  it does not preclude the fact that the SN could be detected
  serendipitously while observing the galaxy, or via Sky Surveys. One
  such is the XMM Serendipitous Source Catalogue, the third release of
  which was in July 2013. This lists sources found by XMM during their
  sky survey. The median flux of all sources is ~2 $\times 10^{-14}$
  erg cm$^2$ s$^{-1}$. We have looked through the list, and notably,
  we have not found a single late time Type IIP SN seen, as would be
  expected if such high density media were present around them. We
  note that there are many Chandra observations of galaxies covering
  the region where optical Type IIP SNe are present, and again there
  is no recorded late time detection of optically observed Type IIps.}

From multiple arguments, it seems likely that at epochs $> 100$ days,
if high mass-loss rate Type IIps were to exist, they should have been
detected.

One final aspect that needs to be taken into account is Comptonization
of the high energy electrons \citep{ci12,ppl13}. Comptonization limits
the maximum energy of escaping photons to $E_{max} \sim
511/{\tau}_{es}^2$, where ${\tau}_{es}$ is the electron scattering
optical depth. We note that this affects the flux in the {\it Chandra}
band only when ${\tau}_{es} \ga 8$. If we assume that ${\tau}_{es}
\le {\tau}_{E}$ as expected, then Comptonization is not important for
our results.

In summary, we can say that it is unlikely that the X-ray emission
from IIPs with higher mass-loss rates would be absorbed so much as not
to have been detected by current X-ray satellites. Thus absorption of
the X-ray emission cannot be used to cover the fact that no X-ray
bright Type IIP SNe are seen.

\section{Discussion}
The two main results from the above sections are that (1) RSGs with
high mass-loss rates ($> 10^{-5} \msun$ yr$^{-1}$) certainly exist in
our galaxy and in other nearby galaxies, and (2) if Type IIP SNe were
to explode in a medium formed by RSGs with these high density winds,
then they should be detectable.

This presents a quandary. If Type IIP SNe arise from RSG stars, it is
surprising that only Type IIP SNe from low mass-loss rate RSG stars
are seen. In the absence of other explanations, one is left with the
conclusion that IIps arise only from those RSGs that have a low
mass-loss rate. Since mass-loss rate is directly related to the
initial progenitor mass, this would then suggest that IIp's arise only
from the lower end of the RSG mass distribution.

We can compute the upper limit based on current observational
constraints. Figure 1, combined with equation 1 and Table 1, suggests
a maximum mass-loss rate of 10$^{-5} \msun$ yr$^{-1}$ for the ambient
medium around Type IIps. Although this refers specifically to the
mass-loss rate measured towards the end of the RSG lifetime, we assume
that the mass-loss rate is constant throughout the RSG
lifetime. Various prescriptions of mass-loss differ somewhat, however
a mass-loss rate of 10$^{-5} \msun$ yr$^{-1}$ corresponds to a RSG
luminosity of about 0.5-2 $\times 10^5 L_{\odot}$ \citep{mj11}. The
Geneva mass-loss rate quoted by \citet{mj11}
$$ \dot{M} = 4.7 \times 10^{-6} \,(L/10^5)^{1.7} $$ gives a luminosity
around 1.6 $\times 10^5 L_{\odot}$. If we use this value, and the
relationship between mass and luminosity suggested by \citet{mj11} $$
M \sim 0.14\, L^{0.41}$$ where mass $M$ and luminosity $L$ are both in
terms of solar values, then we get a mass of $M \sim 19 \msun$ as the
maximum mass of the RSG progenitor of a Type IIP SN. In actual fact
most mass-loss rates are lower, so the value could be somewhat lower.

Amazingly, given the uncertainties, this value is close to others that
have been quoted in the literature. In their paper describing the RSG
problem, \citet{smarttetal09} suggest that the maximum mass of a RSG
progenitor that can give rise to a Type IIP SN is 16.5 $\pm 1.5
\msun$. This is an observational limit derived from the progenitors of
Type IIP SNe that have been carefully identified in pre-explosion
images, with a significance of about 2.4$\sigma$. On the other hand,
recent theoretical work \citep{georgyetal12, grohetal13, georgyetal13}
suggests that rotation restricts the maximum mass of RSGs that explode
to give Type IIP SNe to about 16.8 $\msun$; RSGs larger than this
evolve to become nitrogen-rich (WN) stars before they explode,
resulting presumably in Ib/c SNe. Non-rotating RSGs that become Type
IIP SNe are also limited to an initial mass of 19 $\msun$ due to
increased mass-loss. These stars will not end their lives to produce
Type IIP SNe.

It is interesting that the maximum mass of Type IIP SNe progenitors,
obtained from mass-loss rate limits calculated from X-ray
observations, agree quite well with those obtained by other means. We
must caution however that none of the observational methods are
entirely convincing due to low statistics. In this context it is
interesting to look at the progenitor masses deduced for various Type
IIPs as in Table 1. Those which have mass-loss rates determined all
(except one) fall below the 19 $\msun$ limit calculated herein. This
is not surprising, as the initial progenitor mass, notwithstanding how
it is computed, is somehow related to the derived mass-loss rate, and
therefore would be expected to give a lower initial mass
progenitor. Recently, approaches involving the modelling of late-time
spectra \citep{jerkstrandetal12} also appear to find masses at the
lower end of the RSG mass range, consistent with those from direct
detection of progenitors. These are however in direct contrast to
masses derived from hydrodynamic modelling \citep{ucp07, uc08, uc09,
  ucb10, uc11, uc13}, which tend to be consistently higher. Although
the reasons for this are not known, some that have been speculated are
the neglect of multi-dimensional effects, explosion asymmetries, and
the use of non-evolutionary models in hydrodynamic modelling. This is
not a problem relegated to one particular code or group either;
\citet{tomasellaetal13} found that their best estimate of the
progenitor mass of SN 2012A from their hydrodynamic modelling was 30\%
higher than that from direct mass estimates. It is important that the
results for this discrepancy be understood and evaluated in future,
because the hydrodynamical masses seem to consistently exceed the
upper mass-limit found from other means.

An exception to all the other Type IIP SNe is SN 2009kf, whose
progenitor mass calculated from hydrodynamical modelling (36 $\msun$)
is larger than the maximum mass of stars known to end their lives in
the RSG phase. Although by its light curve it resembles Type IIP, the
explosion energy requirements ($\ga 10^{52} $ ergs) suggest that it is
different from most (all?)  other IIps. The large explosion energy
required to explain its light curve prompted \citet{ucb10} to suggest
that SN 2009kf was caused by the same ``engine'' that leads to the
so-called hypernova explosions. They further speculate that binary
evolution of two massive stars could result in a SN 2009kf type
scenario. \citet{po11} explore fallback accretion onto newly-born
magnetars, and find that for stars with a $> 10 \msun$ H envelope,
this would give a bright Type IIP SNe with a high plateau
luminosity. 

In any realistic scenario, it is clear that SN 2009kf differs from
others in the IIP category. This does not necessarily contradict the
mass-limit found above.  It may simply be an artifact of an inadequate
classification scheme that classifies SNe based on one particular
aspect of their light curve, and expects all SNe that show similar
light curves (in that one aspect) to have similar progenitors. The
lack of a single progenitor for a similar group of SNe has been
mentioned in the context of Type IIn SNe, which show narrow lines on
top of a broad base in their spectra. For many years IIns were thought
to arise from high-mass progenitors, but studies of their host
environments do not seem to confirm this hypothesis \citep{aj08,
  kk12}. There seem to be several indirect indications that multiple
progenitor channels may be involved \citep{dwarkadas11b,
  taddiaetal13}

\section{Summary and Conclusions}
If the X-ray emission from young SNe arises due to thermal
bremsstrahlung, then it should be a strong function of the ambient
density. If we group the light curves by type of SN, those interacting
with the higher density material should have higher luminosities. In
this context it is unusual that a plot of X-ray lightcurves shows that
Type IIP SNe have the lowest X-ray luminosities. This is
counter-intuitive given that IIPs are supposed to arise from RSGs,
which have slow, high mass-loss rate winds and are therefore expected
to have a high-density medium around them. Indeed, measurements of the
medium around some RSGs do suggest that they have wind mass-loss rates
in excess of 10$^{-5} \msun$ yr$^{-1}$. However these are not
reflected in the X-ray lightcurves of the corresponding Type IIP SNe.

In this paper we have examined the reasons for this. It is unlikely
that IIPs do not arise from RSGs; a large envelope is required to
explain the plateau in the light curve. We have examined whether the
X-ray flux from high mass-loss rate IIps could be absorbed by the
surrounding medium. Although this is a complicated questions that
requires investigating several different aspects, the conclusion seems
to be that we should have seen at least some type IIPs exploding in a
higher density medium, either at early (first 100 days) or late
times. The fact that we have not seen any suggests that maybe Type
IIps do not explode in a high mass-loss rate medium ($ > 10^{-5}
\msun$ yr$^{-1}$). If this is true, then it implies a limit of below
19 $\msun$ for the initial mass of a Type IIP progenitor. This is
close to the numbers quoted by direct detection of IIP progenitors, as
well as recent theoretical arguments. This would require changes in
the way we view the evolution of RSG stars with initial masses between
about $\sim$ 19-25 $\msun$.

Further statistics are required before these claims can be
solidified. Even a single solid detection of a Type IIP with a
luminosity exceeding about 10$^{40}$ ergs s$^{-1}$ would create
problems for this assertion. Till then, it appears that different
indicators appear to converge on the suggestion that not all RSGs give
rise to Type IIP SNe, but that those with initial mass $> 17-19 \msun$
may evolve to some other type of star before explosion, perhaps the WN
stars. Observations of more IIPs in the X-ray regime would certainly
help to solidify this claim from the X-ray point of view.

\section*{Acknowledgments}
VVD's research is supported by several Chandra grants. We thank the
anonymous referee for comments that helped to considerably improve the
paper, and R. Chevalier for extremely helpful comments on an earlier
version of this manuscript.

\bibliographystyle{mn2e} \bibliography{paper}

%% This figure uses \includegraphics to scale and rotate the still frame
%% for an mpeg animation.

\bsp

\label{lastpage}

\end{document}